\documentclass{myaa}

\usepackage[varg]{txfonts}

\usepackage{natbib}
\usepackage{graphicx}
\usepackage{epstopdf}
\usepackage{amssymb}
\usepackage{float}

\bibpunct{(}{)}{;}{a}{}{,}

\begin{document}

\title{Theoretical models of the protostellar disks of AS 209 and HL Tau presently forming in-situ planets}

\author{Dimitris M. Christodoulou\inst{1,2}  
\and 
Demosthenes Kazanas\inst{3}
}

\institute{
Lowell Center for Space Science and Technology, University of Massachusetts Lowell, Lowell, MA, 01854, USA.\\
\and
Dept. of Mathematical Sciences, Univ. of Massachusetts Lowell, 
Lowell, MA, 01854, USA. \\ E-mail: dimitris\_christodoulou@uml.edu\\
\and
NASA/GSFC, Laboratory for High-Energy Astrophysics, Code 663, Greenbelt, MD 20771, USA. \\ E-mail: demos.kazanas@nasa.gov \\
}


\def\gsim{\mathrel{\raise.5ex\hbox{$>$}\mkern-14mu
                \lower0.6ex\hbox{$\sim$}}}

\def\lsim{\mathrel{\raise.3ex\hbox{$<$}\mkern-14mu
               \lower0.6ex\hbox{$\sim$}}}

\abstract{
We fit an isothermal oscillatory density model to two ALMA/DSHARP-observed disks, AS 209 and HL Tau, in which planets have presumably already formed and they are orbiting within the observed seven dark gaps in each system. These large disks are roughly similar to our solar nebula, albeit they exhibit milder radial density profiles and they enjoy lower centrifugal support.  We find power-law density profiles with index $k=0.0$ (radial densities $\rho(R) \propto R^{-1}$) and centrifugal support against self-gravity so small that it guarantees dynamical stability for millions of years of evolution. The scale lengths of the models differ only by a factor of 1.9, but the inner cores of the disks are very different: HL Tau's core is 8.0 times larger and 3.6 times denser than the core of AS 209. This results in four dark gaps having formed within the core of HL Tau, whereas no dark gap is found in the core of AS 209. On the other hand, the Jeans frequencies and the angular velocities of the cores are comparable to within factors of 1.9 and 1.6, respectively.}


\authorrunning{ }
\titlerunning{Planet formation in the disks of AS 209 and HL Tau}

\maketitle

\section{Introduction}\label{intro}

In previous work \citep{chr19a}, we presented isothermal models of the solar nebula capable of forming protoplanets long before the protosun is actually formed by accretion processes. This entirely new ``bottom-up'' formation scenario is currently observed in real time by the latest high-resolution ($\sim$1-5~AU) observations of many protostellar disks by the ALMA telescope \citep{alm15,and16,rua17,lee17,lee18,mac18,ave18,cla18,kep18,guz18,ise18,zha18,dul18,fav18,har18,hua18,per18,kud18,lon18,pin18,vdm19}.   In this work, we apply the same theoretical model to the observed disks of AS 209 and HL Tau, two young systems resolved by ALMA/DSHARP observations \citep{alm15,hua18}. We assume that the observations have captured all forming protoplanets and there are no other protoplanets orbiting inside the observed bright rings. That could be proven wrong by future observations, so the present models should be considered as preliminary models based entirely on the current state-of-the-art observations.

An inspection of the ALMA/DSHARP brightness profiles gives the impression that the disks of these systems appear to be similar in structure \citep{alm15,hua18}. They both extend to more than 100 AU, and each disk exhibits seven pronounced dark gaps or annuli. The goal of our modeling effort is to quantify the physical properties of these two disks and to search for differences, based on the arrangements of their dark gaps that are widely believed to already host orbiting protoplanets. The models show only two striking differences, larger than a factor of 2.9 (their $\beta_0$ ratio): the inner core radius of HL Tau is 8.0 times larger than that of AS 209, and the central density of HL Tau is 3.6 times larger than that of AS 209.

The analytic (intrinsic) and numerical (oscillatory) solutions of the isothermal Lane-Emden equation \citep{lan69,emd07} with differential rotation, and the resulting model of the midplane of the gaseous disk have been described in detail in \cite{chr19a} for the solar nebula. Here, we apply in \S~\ref{models2} the same model to the dark gaps of AS 209  and HL Tau, and we compare the best-fit results in these two cases. In \S~\ref{disc}, we summarize our results.

\section{Physical Models of the AS 209 and HL Tau Protostellar Disks}\label{models2}

The numerical integrations that produce oscillatory density profiles were performed with the \textsc{Matlab} {\tt ode15s} integrator \citep{sha97,sha99} and the optimization used the Nelder-Mead simplex algorithm as implemented by \cite{lag98}. This method (\textsc{Matlab} routine {\tt fminsearch}) does not use any numerical or analytical gradients in its search procedure which makes it extremely stable numerically, albeit somewhat slow.

\subsection{Best-Fit models of AS 209 and HL Tau}\label{model1}

\begin{figure}
\begin{center}
    \leavevmode
      \includegraphics[trim=0.2 0.2cm 0.2 0.2cm, clip, angle=0, width=10 cm]{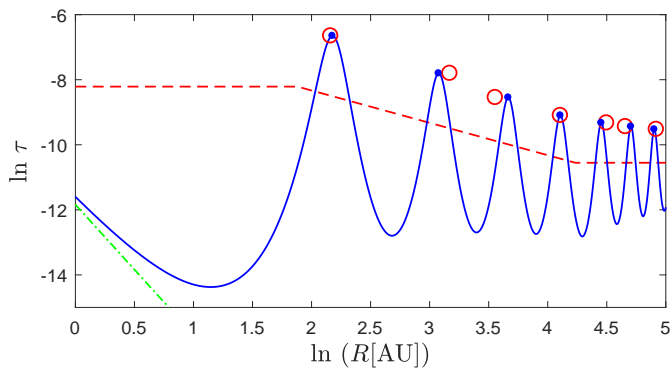}
      \caption{Equilibrium density profile for the midplane of AS 209 disk that has already formed at least seven annular dark gaps (presumably protoplanets) \citep{hua18}. The best-fit parameters are $k=0.0$, $\beta_0=0.0165$, $R_1=6.555$~AU, and $R_2=68.96$~AU. The radial scale length of the disk is only $R_0=0.01835$~AU. The Cauchy solution (solid line) has been fitted to the dark gaps of AS 209 (Table~\ref{table1}) so that its density maxima (dots) correspond to the observed orbits of the protoplanets (open circles). The density maximum corresponding to the location of the fourth (middle) maximum was scaled to a distance of $R_4=60.8$~AU. The mean relative error of the fit is 7.2\%, most of it coming from gaps D24 and D35 (Table~\ref{table1}). The intrinsic analytical solution (dashed line) and the nonrotating analytical solution (dash-dotted line) are also shown for reference. 
\label{fig1}}
  \end{center}
\end{figure}

\begin{figure}
\begin{center}
    \leavevmode
      \includegraphics[trim=0.2 0.2cm 0.2 0.2cm, clip, angle=0, width=10 cm]{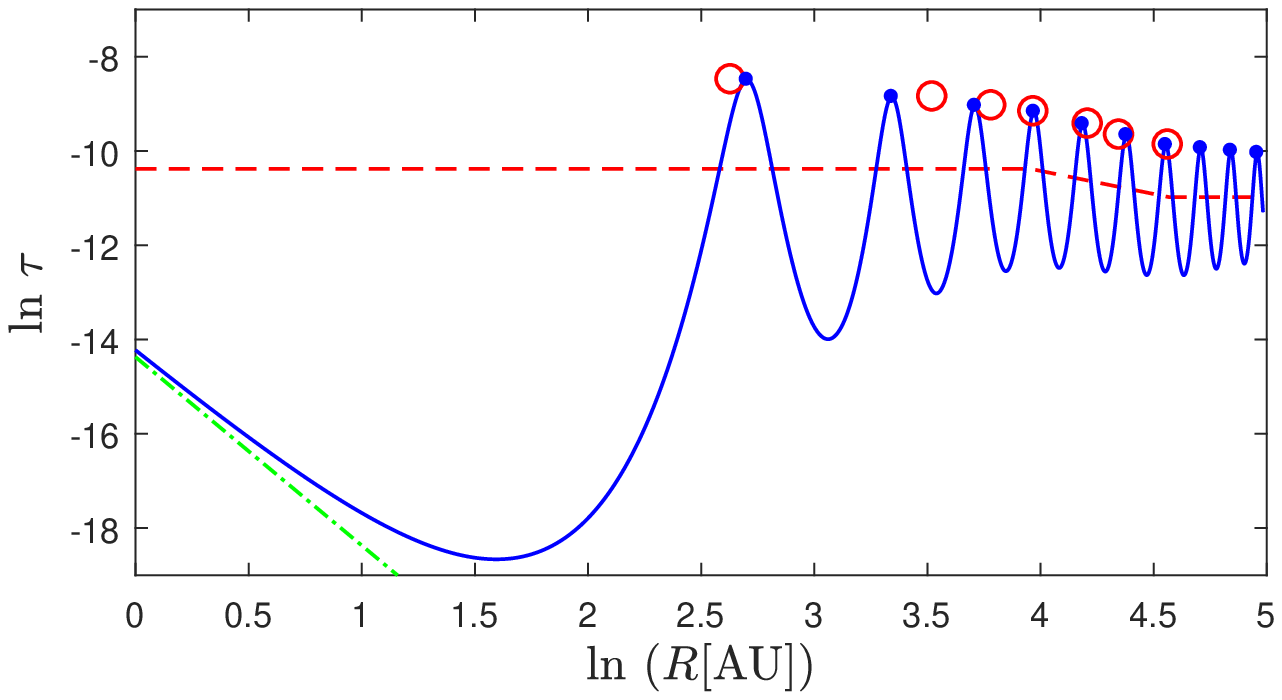}
      \caption{Equilibrium density profile for the midplane of HL Tau disk that has already formed at least seven annular dark gaps (presumably protoplanets) \citep{hua18}. The best-fit parameters are $k=0.0$, $\beta_0=0.00557$, $R_1=52.20$~AU, and $R_2=96.48$~AU. The radial scale length of the disk is only $R_0=0.009730$~AU. The Cauchy solution (solid line) has been fitted to the dark gaps of AS 209 (Table~\ref{table1}) so that its density maxima (dots) correspond to the observed orbits of the protoplanets (open circles). The density maximum corresponding to the location of the fourth (middle) maximum was scaled to a distance of $R_4=53.0$~AU. The mean relative error of the fit is 8.9\%, most of it coming from gaps D34 and D44 (Table~\ref{table1}). The intrinsic analytical solution (dashed line) and the nonrotating analytical solution (dash-dotted line) are also shown for reference. 
\label{fig2}}
  \end{center}
\end{figure}

The radii of the seven dark gaps in both protostellar systems are shown in Table~\ref{table1}. In Figures~\ref{fig1} and~\ref{fig2}, we show the best optimized fits to the dark gaps of AS 209 and HL Tau, respectively. In these models, we have used all four available free parameters ($k$, $\beta_0$, $R_1$, and $R_2$). The mean relative errors are 7.2\% and 8.9\%, respectively, and they all come from the second and third dark gaps in both cases. Because of this discrepancy, we suspect that there may be more dark gaps, yet undetected, in the inner regions of these disks.

The physical properties of these best-fit models are listed in Table~\ref{table2}. We can see from this table several similarities and just two differences between the two disks: The inner core of HL Tau appears to be 8.0 times larger and 3.6 times denser than the disk of AS 209. Furthermore, the centrifugal support in these models is so low, that it practically guarantees their dynamical stability to nonaxisymmetric self-gravitating instabilities \citep[the critical value for the onset of dynamical instabilities is $\beta_*\simeq 0.50$;][]{chr95}.

The outer flat-density regions starting at radius $R_2$ are roughly comparable between the two systems ($\simeq$70 AU and 96 AU, for AS 209 and HL Tau, respectively). This parameter was used in an attempt to obtain a better fit to the outer dark gaps in each system. When we discarded parameter $R_2$, we found significantly worse best-fits for the two systems.

\begin{table}
\caption{Radii of dark gaps in AS 209 and HL Tau \citep[from Table 1 of][]{hua18}}
\label{table1}
\begin{tabular}{ll|ll}
\hline
Gap    & AS 209     & Gap     &  HL Tau \\
Name & $R~(AU)$ & Name  &  $R~(AU)$ \\
\hline
D9   &   08.69   &   D14   &   13.9 \\
D24  &  23.84  &   D34   &   33.9 \\
D35  &  35.04  &   D44   &   44 \\
D61  &  60.8    &   D53   &   53 \\
D90  &  89.9    &   D67   &   67.4 \\
D105 &  105.5  &   D77   &   77.4 \\
D137 &  137     &   D96   &   96 \\
\hline
\end{tabular}
\end{table}

\begin{table*}
\caption{Comparison of the protostellar disk models of AS 209 and HL Tau}
\label{table2}
\begin{tabular}{llll}
\hline
Property & Property & AS 209 & HL Tau \\
Name     & Symbol (Unit) & Best-Fit Model & Best-Fit Model \\
\hline
Density power-law index & $k$                                          &   $0.0$  	     & $0.0$    \\
Rotational parameter & $\beta_0$                                &    0.0165	       &  0.00562   \\
Inner core radius & $R_1$ (AU)                              &   6.555  	       &  52.04    \\
Outer flat-density radius & $R_2$ (AU)                              &   68.96        	   &  90.55   \\
Scale length & $R_0$ (AU)                               &       0.01835    	   &  0.009813   \\
Equation of state & $c_0^2/\rho_0$ (${\rm cm}^5 {\rm ~g}^{-1} {\rm ~s}^{-2}$) & $6.32\times 10^{16}$ & $1.81\times 10^{16}$    \\
Minimum core density for $T=10$~K, $\overline{\mu} = 2.34$ & $\rho_0$ (g~cm$^{-3}$)         &    $5.62\times 10^{-9}$   			&   $1.97\times 10^{-8}$   \\
Isothermal sound speed for $T=10$~K, $\overline{\mu} = 2.34$ & $c_0$ (m~s$^{-1}$) & 188 & 188 \\
Jeans gravitational frequency & $\Omega_J$ (rad~s$^{-1}$)    &    $4.9\times 10^{-8}$ & $9.1\times 10^{-8}$    \\
Core angular velocity & $\Omega_0$ (rad~s$^{-1}$)    &    $8.0\times 10^{-10}$ 	& $5.1\times 10^{-10}$    \\
Core rotation period & $P_0$ (yr)                                 &    249 	   			&  390   \\
Maximum disk size & $R_{\rm max}$ (AU)                &    144 	   			&   102  \\
\hline
\end{tabular}
\end{table*}

\subsection{Comparison between the models of AS 209 and HL Tau}\label{comp}

We show a comparison between the physical parameters of the best-fit models of AS 209 and HL Tau in Table~\ref{table2}. Obviously, these two protoplanetary disks are similar in many of their physical properties. In particular, the power-law index $k=0.0$ is the same, and most physical parameters are to within factors of 1.0-2.9.  The larger differences in $\rho_0$ and $R_1$ were already mentioned in \S~\ref{model1} above. Another striking difference stems from the discrepancy in $R_1$ values: the small core region of AS 209 ($\simeq$6.6 AU) is empty (Fig.~\ref{fig1}), whereas the enormous core of HL Tau ($\simeq$52 AU) hosts four of the seven observed dark gaps (Fig.~\ref{fig2}).

The power-law index $k=0.0$ found in both models was unexpected. It implies a roughly uniform surface density profile for these two disks and a {\it volume density profile} of the form $\rho(R)\propto R^{-1}$, unlike those observed so far in protostellar systems \citep{and07,hun10,lee18}. This parameter may certainly change substantially if more planets will be discovered in these disks by higher resolution observations.

\section{Summary}\label{disc}

In \S~\ref{models2}, we presented our best-fit isothermal differentially-rotating protostellar models of two young systems, AS 209 and HL Tau, recently observed by ALMA/DSHARP \citep{alm15,hua18,guz18,zha18}. These models show seven dark gaps each (Table~\ref{table1}), and it is widely believed that protoplanets have already formed and curved out these dark gaps in the observed otherwise monotonic surface brightness profiles of the disks. The best-fit models are depicted in Figures~\ref{fig1} and~\ref{fig2}, respectively, and a comparison of their physical properties is shown in Table~\ref{table2}.

The two disks appear to be similar in the ALMA/DSHARP observations \citep[e.g.,][]{alm15,hua18}. We found only two pronounced differences in their physical properties: HL Tau's inner core is 8.0 times larger and 3.6 times denser than the core of AS 209. 

The orbital periods of the cores turn out to be 249 yr and 390 yr for AS 209 and HL Tau, respectively. In our solar system, these values fall roughly within the Kuiper belt \citep{tru03,bro04}. Furthermore, both models appear to be extremely stable and long-lived (their $\beta_0$ values are extremely small), so we believe that AS 209 and HL Tau will continue to evolve in a nonviolent fashion (i.e., they will be subject to no planet-planet resonant interactions, no planet-gas disk interactions, and no orbital migrations), allowing for the observed well-organized planet formation to conclude within the observed dark gaps. These results continue to support strongly a ``bottom-up'' scenario in which planets form first in $< 0.1$ Myr \citep{gre10,har18}, followed by the formation of the central star after the gaseous disk has cleared out by accretion and dispersal processes.

\label{lastpage}

\end{document}